\documentclass[11pt,a4paper,fleqn]{article} 
\catcode`\@=11
\@addtoreset{equation}{section}
\def\theequation{\arabic{section}.\arabic{equation}}
\def\appendix{\renewcommand{\thesection}{\Alph{section}}\setcounter{section}{0}
              \renewcommand{\theequation}
            {\mbox{\Alph{section}.\arabic{equation}}}\setcounter{equation}{0}}
\def\maketitle{\thispagestyle{empty}\setcounter{page}0\newpage
                \renewcommand{\thefootnote}{\arabic{footnote}}
                  \setcounter{footnote}0}
\renewcommand{\thanks}[1]{\renewcommand{\thefootnote}{\fnsymbol{footnote}}
               \footnote{#1}\renewcommand{\thefootnote}{\arabic{footnote}}}

\renewcommand{\title}[1]{\begin{center}\Large\bf #1\end{center}\rm\par\bigskip}
\renewcommand{\author}[1]{\begin{center}\Large #1\end{center}}
\newcommand{\address}[1]{\begin{center}\large #1\end{center}}
\newcommand{\pacs}[1]{\smallskip\noindent{\sl PACS numbers:
                       \hspace{0.3cm}#1}\par\bigskip\rm}
\def\babs{\hrule\par\begin{description}\item{Abstract: }\it} 
\def\eabs{\par\end{description}\hrule\par\medskip\rm}
\renewcommand{\date}[1]{\par\bigskip\par\sl\hfill #1\par\medskip\par\rm}

\def\dinfn{Dipartimento di Fisica, Universit\`a di Trento\\ 
                           and Istituto Nazionale di Fisica Nucleare,\\
                                   Gruppo Collegato di Trento, Italia \medskip}

\newcommand{\guido}{Guido Cognola\thanks{e-mail: \sl cognola@science.unitn.it\rm}}
\newcommand{\sergio}{Sergio Zerbini\thanks{e-mail: \sl zerbini@science.unitn.it\rm}}

\def\hs{\qquad}               %%%  horizontal space
\def\nn{\nonumber}            %%%  no number for eqnarray
\def\beq{\begin{eqnarray}}    %%%  begequation/eqnarray
\def\eeq{\end{eqnarray}}      %%%  endequation/eqnarray
\def\ap{\left.}               %%%  open bracket
\def\at{\left(}               %%%  open (
\def\aq{\left[}               %%%  open [
\def\ag{\left\{}              %%%  open {
\def\cp{\right.}              %%%  close bracket
\def\ct{\right)}              %%%  close )
\def\cq{\right]}              %%%  close ]
\def\cg{\right\}}             %%%  close }
\def\R{{\hbox{{\rm I}\kern-.2em\hbox{\rm R}}}}   %%% real numbers
\def\H{{\hbox{{\rm I}\kern-.2em\hbox{\rm H}}}}   %%% Hilbert space
\def\N{{\hbox{{\rm I}\kern-.2em\hbox{\rm N}}}}   %%% natural numbers
\def\C{{\ \hbox{{\rm I}\kern-.6em\hbox{\bf C}}}} %%% complex numbers
\def\Z{{\hbox{{\rm Z}\kern-.4em\hbox{\rm Z}}}}   %%% integers numbers
\def\ii{\infty}                                  %%% infinite
\def\X{\times\,}                                 %%% times
\def\Tr{\mathop{\rm Tr}\nolimits}                  %%% Trace
\renewcommand{\Re}{\mathop{\rm Re}\nolimits}       %%% Real 
\def\dir{/\kern-.7em D\,}                            %%% Dirac Operator
\def\lap{\Delta\,}                                 %%% Laplacian
\def\al{\alpha}
\def\be{\beta}
\def\ga{\gamma}
\def\de{\delta}
\def\ep{\varepsilon}
\def\ze{\zeta}

\def\ka{\kappa}
\def\la{\lambda}

\def\si{\sigma}
\def\om{\omega}
\def\ph{\varphi}

\def\Ga{\Gamma}

\def\La{\Lambda}
\def\Si{\Sigma}

\begin{document}
%\tableofcontents       %%%%%%   index of section

%\preprint{UTF }

\title{Effective action for scalar fields \\ and generalised zeta-function 
regularisation}

\author{\guido and \sergio}
\address{\dinfn}

\date{September 2003}

\babs
Motivated by the study of quantum fields in a 
Friedman-Robertson-Walker (FRW) spacetime, 
the one-loop effective action for a scalar field defined in the ultrastatic
manifold $R\times H^3/\Gamma$, $H^3/\Gamma$ being the finite volume, 
non-compact, hyperbolic spatial section,  is investigated by
a generalisation of  zeta-function regularisation.
It is shown that additional divergences may appear at one-loop level.
The one-loop renormalisability of the model is discussed and making use of 
a generalisation of zeta-function regularisation, the one-loop
renormalisation group equations are derived.
\eabs

%\noindent Keywords: Zeta function-regularisation, functional determinants, topologically non trival spacetimes.
%\newpage

\section{Introduction}
\label{Form}

Within the so-called one-loop approximation in quantum 
field theory, the Euclidean one-loop effective action   may be 
expressed in terms of the sum of the classical action and a contribution depending on a functional determinant of an elliptic differential operator, the so 
called fluctuation operator. 
The ultraviolet one-loop divergences which are eventually present, 
have to be regularised by means of a suitable technique
(for recent  reviews, see \cite{eliz94b,byts96-266-1,byts03b}).
In general, one works on a Euclidean version of the spacetime and 
deals with a self-adjoint, non-negative, second-order differential operator
of the form
\beq
L=-\lap+M^2\,, 
\eeq
where $\lap$ is the Laplace-Beltrami operator and $M^2$ a potential
term depending on the classical (constant) background solution $\phi_c$ and 
in general containing the mass, 
the non-minimal coupling with the gravitational field and a 
possible self-interaction term.

Within the one-loop approximation, one usually splits the original
field $\phi$ in two parts:
the classical background $\phi_c$ and a quantum fluctuation $\Phi$. 
As a result, the theory can be conveniently described by the (Euclidean) 
one-loop partition function
\beq
Z[\phi]=e^{-S[\phi_c]}\int D\Phi\:\:e^{-\int\,dV\,\Phi L\Phi}=e^{-W[\phi]}\,. 
\eeq
Here  $S\equiv S[\phi_c]$ is the classical action, while 
$W\equiv W[\Phi]$ is the one-loop effective action,
which can be related to the determinant of the field operator $L$ by 
\beq
W=-\ln Z=S+\frac{1}{2}\ln\det\frac{L}{\mu^2}\,,
\label{I}
\eeq
$\mu^2$ being a renormalisation parameter, 
which appears for dimensional reasons.

The functional determinant is formally divergent, thus,
the last term in the latter equation  must be 
regularised and in  order to study the one-loop divergences, it is
convenient to make use of a variant of the zeta-function regularisation
\cite{z1,z2,dowk76-13-3224}. To this aim, we select the regularisation 
function \cite{byts96-266-1}
\beq
\rho(\varepsilon,t)=\frac{t^\varepsilon}{\Gamma(1+\varepsilon)}\,,
\eeq
which goes to one as soon as the regularisation parameter 
$\varepsilon$ goes to zero. Then we may write
\beq
W(\varepsilon)=S-\frac{1}{2}\int_0^\ii dt\  
\frac{t^{\varepsilon-1}}{\Gamma(1+\varepsilon)} 
\Tr e^{-tL/\mu^2}=S-\frac{1}{2\varepsilon}\ze(\varepsilon|L/\mu^2)\,,
\label{bb}
\eeq
where, as usual, for the elliptic operator $L$, 
the zeta function is defined by means of the Mellin-like transform 
\beq
\zeta(s|L)=\frac{1}{\Ga(s)}\int_0^\ii dt\ t^{s-1} \Tr e^{-tL}\,,
\hs\hs \ze(s|L/\mu^2)=\mu^{2s}\ze(s|L)\:.
\label{mt}\eeq
For a second order differential operator in 4-dimensions, 
the integral is convergent for $\Re s>2$.

We see that the heat-kernel trace $\Tr e^{-tL}$ plays a preeminent role
in the investigation of the analytical properties of the zeta
function. 
In fact, for a second-order, elliptic non negative operator $L$ in a
boundaryless smooth manifold, one has the short-$t$ asymptotic
expansion
\beq
\Tr e^{-tL}\simeq\sum_{j=0}^\ii A_j(L) t^{j-2}\:,
\label{tas00}
\eeq
where $A_j(L)$ are the well known Seeley-DeWitt coefficients 
\cite{dewi65b,seel67-10-172}. 
As a consequence, $\zeta(s|L)$ is regular at the origin and one gets
the well known result $\zeta(0|L)=A_2(L)$. 
This quantity is easily computable (see for example the recent works 
\cite{kirsten00,vass03}) and depends only on coupling constants 
and geometrical invariants.

By expanding (\ref{bb}) in Taylor's series one obtains
\beq
W(\ep)&=&S-\frac{\zeta(0|L/\mu^2)}{2\ep}-\frac{\zeta'(0|L/\mu^2)}{2}
+O(\ep)
\label{dow33}
\eeq
and the regularised one-loop effective action $W_R$ can be  defined
by taking the finite part of $W(\ep)$ in the limit $\ep\to0$,
that is  
\beq 
W_R=S-\frac{\zeta'(0|L/\mu^2)}2=S-\frac{\zeta'(0|L)}2
-\ln\mu^2\:\frac{\ze(0|L)}2\:. 
\eeq
This leads to the usual zeta-function regularisation 
prescription \cite{z2}, i.e. 
\beq 
\ln\det L=-\ze'(0|L).
\eeq
The one loop-divergences are governed by
$\zeta(0|L)=A_2(L)$, which does not depend on the 
arbitrary scale parameter $\mu$.
The coefficient $A_2(L)$ also determines the beta functions of the 
model, namely its one-loop renormalisation group equations (RGEs) and its 
one-loop renormalisability. 
In fact, the RGEs can be obtained by assuming that all coupling 
constants appearing in the renormalised effective action 
$W_R$ are depending on $\mu$ and requiring 
\beq
\mu\:\frac{d}{d\mu} W_R\equiv\mu\:\frac{d}{d\mu}S
-\frac{\mu}2\:\frac{d}{d\mu}\:\zeta'(0|L/\mu^2) =0\,.
\eeq
In this way, at one-loop level one obtains the renormalisation group equations
in the form \cite{molina,cognola93}
\beq
\mu\:\frac{d}{d\mu}S=\ze(0|L)=A_2(L)\,.
\eeq

In this paper, we would like to discuss a more general case corresponding 
to the presence of  logarithmic terms in the heat-trace asymptotics. 
One may have logarithmic terms in the heat-kernel trace 
in the case of non smooth manifolds, for example when one considers
the Laplace operator on higher dimensional cones \cite{bord96, cogn97}, 
but also in 4-dimensional spacetimes with a 3-dimensional, non-compact,
hyperbolic spatial section of finite volume \cite{byts97}. 
More recently the presence of logarithmic terms in self-interacting
scalar field theory defined on manifolds with non-commutative
coordinates have also  been pointed out \cite{byts01,byts02}.

The content of the paper is the following. 
In Section 2, the heat-kernel asymptotics with logarithmic 
terms are considered and the consequences of their
presence discussed in some detail.  
A generalisation of zeta-function regularisation is 
proposed and the generalised one-loop RGEs are derived. 
In Section 3, an explicit example related to FRW by a conformal 
transformation is investigated in detail, 
and  the generalised RGEs are explicitly  written down. 
The conclusions and two Appendices end the paper.

\section{Heat-kernel asymptotics with logarithmic terms}

In this Section we will discuss the modification of the formalism due to
the presence of logarithmic terms in the heat-kernel asymptotics.
The starting point are Eqs.~(\ref{I}) and (\ref{bb}). We have to discuss the 
meromorphic extension of the zeta function, which depends on the form 
of the heat-kernel asymptotics. 

Let us suppose to deal with a quite general expression of the kind
\beq
\Tr e^{-tL}\simeq\sum_{j=0}^\ii B_j\,
t^{j-2}+\sum_{j=0}^\ii P_j\, \ln t\:
t^{j-2}
\:,\label{tas0}
\eeq
where now, with respect to the standard case, 
new terms containing $\ln t$ are present in the expansion. 

In order to obtain the meromorphic continuation of the zeta function, 
we use the Mellin-like representation, Eq.~(\ref{mt}).
The original integral over $t$ can be split into two integrals, 
the first from 1 to $\ii$, 
which gives analytic contributions to zeta function
and the second from 0 to 1, which gives rise the poles and can be explicitly 
computed using the small $t$ expansion (\ref{tas0}).
In this way we get
\begin{eqnarray}
\zeta(s|L)=\frac{1}{\Gamma(s)}\at \sum_{j=0}^\infty\frac{B_j(L)}{s+j-2}
-\sum_{j=0}^\infty\frac{P_j(L)}{(s+j-2)^2}+J(s) \ct \,,
\label{merom-log}
\end{eqnarray}
the function $J(s)$ being analytic.

We see that in contrast with the standard situation, 
here the zeta function has also double poles and,  
if $P_2$ is non vanishing, it is no longer regular at the origin, 
but it has a simple pole with residue $-P_2$.
Another important consequence for physics is that,
due to the presence of logarithmic terms,
the heat-kernel coefficients $B_n$, with respect to 
scale transformations, transforms in a non homogeneous manner.
This can be easily seen by replacing 
the dimensionless parameter $t$ with $t/\mu^2$ in the heat expansion.
In this way
\beq
\Tr e^{-tL/\mu^2}&\simeq&\sum_{j=0}^\ii B_j(L/\mu^2)\,t^{j-2}
+\sum_{j=0}^\ii P_j(L/\mu^2)\,\ln t\:t^{j-2}
\nn\\
&=&\sum_{j=0}^\ii B_j(L)\,\at\frac{t}{\mu^2}\ct^{j-2}
+\sum_{j=0}^\ii P_j(L)\,\ln\at\frac{t}{\mu^2}\ct
\:\at\frac{t}{\mu^2}\ct^{j-2}
\:,\eeq
from which it follows
\beq 
B_n(L/\mu^2)&=&\mu^{4-2n}\:\aq B_n(L)-\ln\mu^2\:P_n(L)\cq\:,
\nn\\
P_n(L/\mu^2)&=&\mu^{4-2n}\:P_n(L)\:.
\label{A2log}\eeq
In particular, in contrast with the standard case, the coefficient 
$B_2$ is not scale invariant.
It is convenient to split the $B_n$ coefficients in two parts, 
that is $B_n=A_n+Q_n$, where $A_n$ respresent the standard coefficients,
obtained as integral of the local geometric quantities $a_n$,
namely
\beq 
A_n(L)=\frac{1}{(4\pi)^2}\:\int\:dV\:a_n(x|L)\:,
\hs\hs A_n(L/\mu^2)=\mu^{4-2n}\:A_n(L)\:,
\eeq
while the second part $Q_n$ is strictly connected 
with the presence of logarithmic terms and transforms according to
\beq 
Q_n(L/\mu^2)&=&\mu^{4-2n}\:\aq Q_n(L)-\ln\mu^2\:P_n(L)\cq\:,\nn\\
Q_2(L/\mu^2)&=&Q_2(L)-\ln\mu^2\:P_2(L)\:,\nn\\
P_2(L/\mu^2)&=&P_2(L)\:.
\eeq

The consequences of the presence of such a pole at the origin 
on the one-loop effective action can be investigated by 
using the  regularisation of Section 1,  namely
\beq
\ln\det\at L/\mu^2\ct_\ep=-\int_0^\ii dt\: 
\frac{ t^{\ep-1}}{\Gamma(1+\varepsilon)}\:\Tr e^{-tL/\mu^2}
=-\frac{\zeta(\ep|L/\mu^2)}{\ep}
=-\frac{\omega(\ep|L/\mu^2)}{\ep^2}\,.
\label{regmt}
\eeq
In the latter equation we have conveniently introduced the new 
kind of zeta function $\om$, regular at the origin, by means of 
the relation
\beq
\omega(s|L)=s\zeta(s|L)\,,\hs\hs
\omega(s|L/\mu^2)=s\:\mu^{2s}\ze(s|L)=\mu^{2s}\om(s|L)\,.
\label{nz}
\eeq

We may expand $\omega$ in Taylor's series around $s=0$, 
obtaining in this way
\beq
\ln\det\at L/\mu^2\ct_\ep
=-\frac{1}{\ep^2}\omega(0|L/\mu^2)
-\frac{1}{\ep} \omega'(0|L/\mu^2)
-\frac{1}{2}\omega''(0|L/\mu^2) +O(\ep) \,.
\label{regmt1}
\eeq
As a consequence, the one loop-divergences are governed by the two 
coefficients 
$\omega(0|L/\mu^2)$and $\omega'(0|L/\mu^2)$, 
while the non trivial finite part is given by 
$\frac{1}{2}\omega''(0|L/\mu^2) $. 
This suggests a generalisation of the 
zeta-function regularisation for a functional determinant 
associated with an elliptic operator $L$, namely \cite{byts03b}
\beq
\ln\det L=-\frac{1}{2}\omega''(0|L)\,.
\label{nreg}
\eeq
Of course, this reduces to the usual zeta-function regularisation 
when $\zeta(s|L)$ is regular at the origin.

The two coefficients governing the one-loop divergences can be 
computed making use of the meromorphic structure
\beq 
\om(s|L)=\frac{1}{\Ga(s)}\:
\sum_{j=0}^\infty\frac{s\:B_j}{s+j-2}
-\frac{1}{\Ga(s)}\:\sum_{j=0}^\infty\frac{s\:P_j}{(s+j-2)^2}
+\frac{s}{\Ga(s)}\:J(s)\,.
\label{omMS}\eeq

One has
\beq
\omega(s|L)=-P_2(L)+[B_2(L)-\ga P_2(L)]\:s+O(s^2)
\label{cit}
\eeq
and so, using Eq.~(\ref{nz}) or alternatively Eq.~(\ref{A2log}),
one has 
\beq
\omega(0|L/\mu^2)=-P_2(L)\,,\hs\hs
\omega'(0|L/\mu^2)= B_2(L)-(\gamma+\ln\mu^2)P_2(L)\,,
\label{nea0}
\eeq
$\ga$ being the Euler-Mascheroni constant.

Recall that he model is one-loop renormalisable, if the dependence of 
$B_2$ and $P_2$ on the background field has the same algebric structure of
the classical action and the divergences may be reabsorbed by the redefinition 
of mass and coupling constants.  

With regard to the derivation of one-loop RGEs, they may be obtained again by 
assuming that the mass and all coupling constants 
appearing in the classical action are depending on $\mu$ and 
requiring 
\beq
\mu\:\frac{d}{d\mu}\:W_R=0\,,
\label{o}
\eeq
where now the functional determinant appearing in Eq.~(\ref{I}) 
is regularised according to Eq.~(\ref{nreg}). 
In this way we get
\beq
W_R&=&S-\frac{1}{4}\omega''(0|L/\mu^2) \nn \\
&=&S-\frac{1}{4}\aq
\omega''(0|L)+2\ln\mu^2\omega'(0|L)+(\ln\mu^2)^2\omega(0|L)\cq\,.
\label{xxx}
\eeq
Making use of Eqs.~(\ref{nea0})-(\ref{xxx}), we finally get
(at one-loop level)
\beq
\mu\:\frac{d}{d\mu}\:S=\omega'(0|L)+\ln\mu^2\omega(0|L)=
B_2(L)-\at\gamma+\ln\mu^2\ct\:P_2(L) \,.
\label{nreg1}
\eeq 
If the theory is renormalisable, the action and the 
heat coefficients have the same structure in terms of the fields. 
More precisely, if the action has the form
\beq 
S=\int\:dV\:\sum_\al\,\la_\al(\mu)\:F_\al\:,
\label{az1}\eeq
then 
\beq 
B_2(L)=\int\:dV\:\sum_\al\,k_\al(\mu)\:F_\al\:,\hs\hs
P_2(L)=\int\:dV\:\sum_\al\,h_\al(\mu)\:F_\al\:,
\label{az2}\eeq
where $F_\al\equiv(1,\phi^2/2,\phi^4/24,...)$ are the independent
building blocks, 
$\la_\al\equiv(\La,m^2,\la,...)$ the
collection of all coupling constants, including the ones 
concerning the gravitational action, while $k_\al$ and $h_\al$
are constants, which can be directly read off from the form of the heat
coefficients.

From Eqs.~(\ref{nreg1})-(\ref{az2}) one obtains the differential equations
for the beta functions in the form
\beq 
\be_\al\equiv\mu\:\frac{d\la_\al}{d\mu}=
k_\al-(\ga+\ln\mu^2)\:h_\al\:,
\label{betaF}\eeq
which of course give the usual result when $P_2=0$.

\section{Scalar fields in a Friedman-Robertson-Walker spacetime}

Now we will provide an application 
of the formalism previously developed.
We shall study a scalar field defined on a spacetime 
of the kind $\R\times\Sigma_3$, where $\Sigma_3$ is the 
constant curvature spatial section. 
The physical motivations are due to the fact that, in a suitable coordinates
system, the metric of the FRW spacetime is conformally related to the metric 
of $R\times\Sigma_3$ and moreover, as we shall see, 
in usual cases, the renormalisation properties do not depend on the
conformal transformation. This statement will be  clarified later on. 

We start with a 4-dimensional FRW spacetime with the standard metric 
\beq
ds^2=-dT^2+a^2(T)\:d\sigma_3^2\,,
\label{h}
\eeq
$d\sigma_3^2$ being the metric associated with the 
3-dimensional manifold with constant curvature.
Then we introduce the related conformal time $\eta$ by
\beq
\eta=\int\frac{dT}{a(T)}\,.   
\eeq
In this way the metric assumes the form
\beq
ds^2=a^2(\eta)\at-d\eta^2+d\sigma_3^2 \ct\,.
\label{hc}
\eeq
This means that (locally) the spacetime is conformally 
related to a constant curvature manifold $M^4=R\times\Sigma_3$, 
possibly equipped with a non trivial topology. 
In particular we shall investigate in detail the case of a non-compact and
non-smooth hyperbolic spatial section $\Si_3=H^3/\Gamma$, with finite volume,  
$H^3$ being the 3-dimensional hyperbolic manifold and $\Gamma$
a discrete group of isometries containing hyperbolic and parabolic 
elements \cite{byts97}. 

We denote by $\tilde M^4$ the original spacetime with the metric
$\tilde g_{ij}$ conformally related to the metric $g_{ij}$
of the constant curvature manifold $M^4$, that is
\begin{eqnarray}
\tilde g_{ij}=e^{2\sigma}g_{ij}\:,\hs\hs
\sigma=\ln a(\eta)\:,
\end{eqnarray}
$\sigma:=\sigma(x)$ being a scalar function.

In general, by conformal transformations, the partition function
is not invariant (see the Appendix A), but one has 
\begin{eqnarray}
\tilde W= W[\tilde\phi,\tilde g]=W[\phi,g]-\ln J[g,\tilde g]
\:, \end{eqnarray}
where $J[g,\tilde g]$ is the Jacobian of the conformal transformation. 
Such a Jacobian (also called cocycle function or induced effective action) 
can be computed for any infinitesimal conformal transformation (see for
example  \cite{brown84,dowker84,gusy,buch} and references cited 
therein). Its expression in 4-dimensions reads (see the Appendix A)
\beq 
\ln J[g,\tilde g]=
\frac{1}{(4\pi)^2}\int_0^1\:dq\int\:d^4x\:\sqrt{g^q}\:\aq
b_2(x|L_q)-(\ga+\ln\mu^2)\:p_2(x|L_q)\cq\:,
\label{lnJ}\eeq
where $b_2(x|L_q)$ and $p_2(x|L_q)$ are the local quantities related 
to the coefficients $B_2(L_q)$ and $P_2(L_q)$ respectively, 
while $L_q$ is the field operator in the metric
$g^q_{ij}=\exp(2q\si)\:g_{ij}$.
Thus, in principle, the knowledge of the partition
function $Z[\phi,g]$ in the manifold $M^4$
and the heat coefficients $b_2$ and $p_2$, 
are sufficient in order to get the
partition function in the original manifold $\tilde M^4$.
For such a reason here we shall study the heat-kernel asymptotics
and the one-loop effective action for scalar fields in
$M^4=\R\X H^3/\Ga$. If $\Ga$ contains parabolic elements,
the heat-kernel asymptotics for the Laplacian
contains also logarithmic terms \cite{byts97}.

As can be trivially seen, in the standard case 
the relevant heat-kernel coefficient  $a_2(x|L/\mu^2)$ 
does not depend on $\mu$ and the Jacobian $J[g,\tilde g]$ is finite.
This means that $W$ and $\tilde W$ have the same
one loop divergences and give rise to the same 
renormalisation group equations.
The situation completely change in the ``non-standard case'' we are 
going to consider, since the Jacobian factor explicitly depeds on $\mu$,
as one can see looking at Eq.~(\ref{lnJ}).

\subsection{Heat-kernel expansion for scalar fields on $\R\X H^3/\Ga$}  

We start with the classical Euclidean action for a massive,
self-interacting scalar field in $M^4\equiv\R\times H^3/\Gamma$,
$H^3$ being the 3-dimensional hyperbolic manifold and 
$\Ga$ a group containing the identity, hyperbolic and parabolic elements. 
$H^3/\Ga$ is a rank-1 symmetric space with constant curvature $R$.
This latter is also the scalar curvature of $M^4$ 
(For more details concerning the geometry of this spatial 
section, see \cite{byts97}). 

The action for the matter field has the form
\begin{eqnarray} S[\phi,g]=\int
\left[-\frac12\phi\Delta\phi+V_c(\phi)\right]\sqrt{g}\:d^4x
\:,\label{Sc}\end{eqnarray} 
where the classical potential reads
\begin{eqnarray} V_c(\phi)=\frac{\lambda\phi^4}{24}+
\frac{m^2\phi^2}2+\frac{\xi R\:\phi^2}2 
\:,\end{eqnarray}
$m$ being the mass of the field,
$\lambda$ the self-interacting coupling constant and
$\xi$ a coupling constant which takes into account of a possible
non minimal coupling between matter and gravitation  
(see, for example \cite{hu83,cognola93}).

Within the one-loop approximation, one splits the field as
$\phi=\phi_c+\Phi$, $\phi_c$ being the background (classical) field and
$\Phi$ the quantum fluctuation. 
In this way, the relevant operator associated with the 
quantum fluctuation is given by
\begin{eqnarray} 
L=-\lap+M^2\:,\hs\hs M^2=m^2+\xi R+\frac{\lambda\phi^2_c}{2}\:.
\end{eqnarray} 
Since we are dealing with an ultrastatic spacetime, we have
\beq
L=-\partial^2_\eta-\lap_3+M^2=-\partial^2_\eta+L_3\:,\hs\hs
L_3=-\lap_3+M^2
\:,\label{nmop}\eeq
$\lap_3$ being the Laplace-Beltrami operator acting on functions 
in $H^3/\Ga$.

In the regularisation scheme we have proposed in previous Sections,
the one-loop effective action is given by
\beq
W=S+\frac{1}{2}\ln\det\frac{L}{\mu^2}=
S-\frac14\:\om''(0|L/\mu^2)\:.\label{ep}\eeq
For the case under consideration, the heat-kernel trace has the form 
\beq
\Tr e^{-tL}=\frac{\ell}{\sqrt{4\pi t}}\Tr e^{-tL_3}\,,
\label{trL4}
\eeq 
$\ell$ being the ``infinite volume'' of $\R$.
For the rank-1 symmetric space $H^3/\Ga$, the trace of the heat-kernel
can be computed by using the Selberg Trace Formula.
In our case (the group $\Ga$ contains identity, 
hyperbolic and parabolic elements) we get (see Ref.~\cite{byts97} for details)
\beq
\Tr e^{-tL_3}\sim \sum_{j=0}^\ii\aq 
B_j(L_3)+P_j(L_3)\:\ln t\cq\:t^{j-3/2}
\:,\label{trL3}
\eeq
where
\beq 
B_0(L_3)&=&\frac{v_F}{(4\pi)^{3/2}}\:,\nn\\
B_1(L_3)&=&-\frac{v_F\:\de^2}{(4\pi)^{3/2}}+\frac{C}{\sqrt{4\pi}}\:,\nn\\
B_2(L_3)&=&\frac{v_F\:\de^4}{2(4\pi)^{3/2}}+\frac{1}{6\sqrt{\pi}}
-\frac{C\:\de^2}{\sqrt{4\pi}}\:,
\eeq
\beq
P_0(L_3)=0\:,\hs
P_1(L_3)=\frac{1}{8\sqrt\pi}\:,\hs
P_2(L_3)=-\frac{\de^2}{8\sqrt\pi}\:,
\eeq
$C$ being a known constant, $v_F$ the (dimensionless) fundamental volume 
and $\de^2=|\ka|+M^2$, where 
$\ka=R/6$ is the negative, constant (Gaussian) curvature of
of the manifold. 
We have written only the coefficients which we need in the paper,
but in principle all coefficients can be computed.
We have also used units in which $\ka=-1$.

From Eqs.~(\ref{trL4}) and (\ref{trL3}) we immediately obtain the
expansion we are interested in, that is 
\beq
\Tr e^{-tL}\sim\sum_{j=0}^{\ii}
\aq B_j(L)+P_j(L)\:\ln t\cq\:t^{j-2}
\:,\label{te}\eeq
where trivially
\beq 
B_j(L)=\frac{\ell\:B_j(L_3)}{\sqrt{4\pi}}\:,\hs\hs
P_j(L)=\frac{\ell\:P_j(L_3)}{\sqrt{4\pi}}\:.
\eeq 
Then, in the standard units, we finally obtain
\beq 
B_0(L)&=&\frac{V}{16\pi^2}\:,\nn\\
B_1(L)&=&-\frac{V}{16\pi^2}\at\de^2+\frac{2\pi\,C\,R}{3v_F}\ct
\:,\nn\\
B_2(L)&=&\frac{V}{16\pi^2}\at\frac{\de^4}{2}+\frac{\pi\,R^2}{27v_F}
+\frac{2\pi\,C\,\de^2\,R}{3v_F}\ct\:,
\eeq
\beq
P_0(L)=0\:,\hs
P_1(L)=-\frac{V}{16\pi^2}\:\frac{\pi\,R}{6v_F}\:,\hs
P_2(L)=\frac{V}{16\pi^2}\:\frac{\pi\,R\,\de^2}{6v_F}\:.
\eeq
Expanding the previous quantity we have in particular
\beq 
B_2(L)&=&\frac{V}{16\pi^2}\ag 
\frac{m^4}{2}+\la m^2\:\frac{\phi_c^2}2+3\la^2\:\frac{\phi_c^4}{24}
+\aq\la\at\xi-\frac16\ct+\frac{2\pi\la\,C}{3v_F}\cq\:\frac{R\,\phi_c^2}2
\cp\nn\\ &&\hs
+\aq m^2\at\xi-\frac16\ct+\frac{2\pi m^2\,C}{3v_F}\cq\:R
\nn\\ && \ap\hs\hs
+\aq\frac{\pi}{27v_F}+\frac12\at\xi-\frac16\ct^2
+\frac{2\pi\,C}{3v_F}\at\xi-\frac16\ct\cq\:R^2\cg\:,
\label{b2}\eeq
\beq
P_2(L)&=&\frac{V}{16\pi^2}\ag
\frac{\pi\la}{6v_F}\:\frac{R\,\phi_c^2}2
+\frac{\pi m^2}{6v_F}\:R
+\frac{\pi}{6v_F}\at\xi-\frac16\ct\:R^2\cg\:.
\label{p2}\eeq

\subsection{The renormalised one-loop effective action and the one-loop 
renormalisation group equations }

Here we would like to explicitly compute the beta functions
for the model considered above.
In order to have the renormalisation of the model, to the
matter action (\ref{Sc}) we have to add the action for the 
(classical) gravitational field, then 
\begin{eqnarray} S[\phi,g]=\int
\left[\La-\frac12\phi\Delta\phi+V_c(\phi)
+g_1\:R+g_2\:R^2\right]\sqrt{g}\:d^4x
\:.\label{Stot}\end{eqnarray}
$\Lambda$ being the bare cosmological constant, 
In this way we have the independent building blocks
\beq
F_\al\equiv\at1,\frac{\phi^2}2,\frac{\phi^4}{24},\frac{R\:\phi^2}2,R,R^2\ct
\eeq
 and the corresponding coupling constants
$\la_\al(\mu)\equiv(\La,m^2,\la,\xi,g_1,g_2)$.

Now, from Eqs.~(\ref{betaF}), (\ref{b2}) and (\ref{p2}) we
directly get
\beq 
\mu\:\frac{d\La}{d\mu}=\frac{m^4}{32\pi^2}\:,\hs\hs
\mu\:\frac{dm^2}{d\mu}=\frac{\la m^2}{16\pi^2}\:,\hs\hs
\mu\:\frac{d\la}{d\mu}=\frac{3\la^2}{16\pi^2}\:,
\label{rge1}\eeq
\beq
\mu\:\frac{d\xi}{d\mu}&=&\frac{\la(\xi-1/6)}{16\pi^2}
+\frac{\la\,C}{24\pi v_F}
-\frac{\la(\ga+\ln\mu^2)}{96\pi v_F}\:,
\nn\\
\mu\:\frac{dg_1}{d\mu}&=&\frac{m^2(\xi-1/6)}{(16\pi^2}
+\frac{m^2\,C}{24\pi v_F}
-\frac{m^2(\ga+\ln\mu^2)}{96\pi v_F}\:,
\nn\\
\mu\:\frac{dg_2}{d\mu}&=&\frac{1}{432\pi v_F}
+\frac{(\xi-1/6)^2}{32\pi^2}
+\frac{C(\xi-1/6)}{24\pi v_F}
-\frac{(\xi-1/6)(\ga+\ln\mu^2)}{96\pi v_F}\:.
\label{rge2}\eeq

Some remarks are in order. First, as in the 4-dimensional smooth case, 
there is no renormalisation of the wave function. 
Second, the RGEs concerning 
the  coupling constants $\Lambda$, $m^2$  and $\lambda$ (see Eq.~(\ref{rge1}))
are exactly the same which one has in the smooth case 
(see for example Refs. \cite{hu83,guido}),
while the RGEs related to $\xi$, $g_1$ and $g_2$ are modified
by the presence of the parabolic elements of the group $\Ga$.

\section{Conclusion}

In this paper we have started an  investigation concerning  the one-loop 
effective action for  a 
scalar field in a 4-dimensional FRW spacetime. The one-loop effective action 
may be considered as the sum of two 
contributions. The first one can be computed
considering the field in a constant curvature spacetime,
conformally related to the original manifold, while the second one
takes its origin in the Jacobian of the conformal transformation. 
This latter contribution is what is usually 
called the conformal induced effective action.

Here we have studied in detail the first contribution in the  particular
spacetime $R \times H^3/\Gamma$.  
Due to the non trivial topology of the manifold we have considered, new
ultraviolet divergences appear in the one loop-effective action. 
At one-loop level,
we have shown that all divergences may reabsorbed by suitable counterterms 
in the classical action. The new divergences  depend  on the coefficient 
\beq
P_2(L)&=&\frac{V}{16\pi^2}\aq
\frac{\pi\la\,R\,\phi_c^2}{12v_F}
+\frac{\pi m^2\,R}{6v_F}
+\frac{\pi(\xi-1/6)\,R^2}{6v_F}\cq\:.
\eeq
In the massless and free case with 
the conformal coupling $\xi=1/6$, one has
$P_2=0$ and thus, in this particular situation the usual evaluation of 
the effective action by means of zeta-function 
methods works without any modification. It has to be noted that 
the choice $m=0$, $\la=0$ and $\xi=1/6$ is consistent with 
RGGs. 
If the $P_2$ coefficient is not vanishing,
a generalisation of the zeta-function techniques has been used and 
the the RGEs have been derived in a consistent way, checking that the model is
indeed renormalisable at one-loop level. 

As far as the anomaly induced effective action is concerned, 
in the case $P_2=0$ it can be computed making use of the general 
expressions reported in Section 1 and in 
Appendix A \cite{reigert,fradkin,buch}. 
The evaluation of it when $P_2\neq0$ is not an 
easy task and we will investigate it in a next future.

\appendix
\section{Conformal Transformations}

In this Appendix we shall consider the conformal properties of the first 
non trivial Seeley-DeWitt coefficients (see also Ref.~\cite{dowker89}).
For more generality, here we work in a $D$-dimensional Euclidean manifold
and for convenience we use the scalar density $\varphi=g^{1/4}\:\phi$ 
in place of the scalar field $\phi$.
The classical action then assumes the form
\beq 
S=\int\:d^Dx\:\sqrt{g}\phi\:L\:\phi
=\int\:d^Dx\:\ph\:L\:\ph\:,
\eeq
where $L=-\lap_g+m^2+\xi R$ is a Laplacian-like operator 
($\lap_g=g^{ij}\nabla_i\nabla_j$, $\nabla_i$ being the covariant 
derivative).

A conformal transformation is defined by
\beq 
\tilde g_{ij}=e^{2\si}\:g_{ij}\:,
\hs\hs\tilde g=|\det\tilde g_{ij}|=e^{D\si}\:g\:,
\hs\hs i,j=0,1,...,D-1\:,
\eeq
\beq 
\tilde\ph=e^{\si}\ph\:,\hs\hs\tilde\phi=e^{(1-D/2)\si}\phi\:,
\eeq
$\si\equiv\si(x)$ being a generic scalar function. 

By a straightforword computation, for the connection coefficients,
the Riemann and Ricci tensors and scalar curvature one obtains 
respectively
\beq 
&& \tilde\Ga^k_{ij}=\Ga^k_{ij}+\Si^k_{ij}\:,
\nn\\
&& \tilde R^i_{jrs}=R^i_{jrs}+\Si^i_{jrs}\:,
\nn\\
&& \tilde R_{ij}=R_{ij}+\Si_{ij}\:,
\nn\\
&& \tilde R=e^{-2\si}(R+\Si)\:
\eeq
where 
\beq
\Si^k_{ij}=\si_i\de^k_j+\si_j\de^k_i-\si^kg_{ij}\:,
\hs\hs\si_k=\partial_k\si\:. 
\eeq
Furthermore, for any scalar function $f$
\beq 
\lap_{\tilde g}f=e^{-2\si}\aq\lap_g+(D-2)\si^k\partial_k\cq\:f\:.
\eeq

In order to make explicit computations, now we introduce 
the useful notation
\beq 
&& \si_{ij}=\nabla_i\nabla_j\si\:,\hs\hs\si^k_k=\lap\si\:,
\nn\\
&& B_{ij}=B_{ji}=\si_{ij}-\si_i\si_j+\frac{g_{ij}}{2}\:\si^k\si_k\:,
\nn\\
&& B=B^k_k=\lap\si+\frac{D-2}{2}\:\si^k\si_k\:,\nn\\
&& B^{ij}B_{ij}=\si^{ij}\si_{ij}-2\si^{ij}\si_i\si_j
-\lap\si\:\si^k\si_k+\frac{D\:(\si^k\si_k)^2}{2}
\:.
\eeq
In this way
\beq 
&& \Si^i_{jrs}=\nabla_r\Si^i_{sj}-\nabla_s\Si^i_{rj}
+\Si^i_{rl}\Si^l_{sj}-\Si^i_{sl}\Si^l_{rj}
=-\aq g_{ir}B_{js}-g_{is}B_{jr}+g_{js}B_{ir}-g_{jr}B_{is}\cq\:,
\nn\\
&&\Si_{ij}=\Si^k_{ikj}=-\aq (D-2)B_{ij}+B^k_k\:g_{ij}\cq\:,
\nn\\
&&\Si=g^{ij}\Si_{ij}=-2(D-1)\lap_g\si-(D-1)(D-2)\si^k\si_k
\nn\\
&&\hs\hs\hs\hs =-2(D-1)B  \Si^{ijrs}\Si_{ijrs}=4(D-2)B^{ij}B_{ij}+4B^2\:,
\nn\\
&& \Si^{ij}\Si_{ij}=(D-2)^2B^{ij}B_{ij}+(3D-4)B^2\:.
\eeq
Now it is  easy to verify that the Weyl tensor 
\beq 
 C_{ijrs}=R_{ijrs}&+&\frac{1}{D-2}\:
   \at g_{ir}R_{js}-g_{is}R_{jr}+g_{js}R_{ir}-g_{jr}R_{is}\ct
\nn\\
&-&\frac{1}{(D-1)(D-2)}\:R\:\at g_{ir}g_{js}-g_{is}g_{jr}\ct\:,
\eeq
is conformally invariant, while the Gauss-Bonnet 
\beq 
G=R^{ijrs}R_{ijrs}-4R^{ij}R_{ij}+R^2\:, 
\eeq
transforms according to
\beq 
\tilde G=G+8(D-3)R^{ij}B_{ij}-4(D-3)RB-4(D-2)(D-3)\at B^{ij}B_{ij}-B^2\ct\:.
\eeq
Recall that in 4-dimensions, $G$ is a total divergence.

By definition, $\tilde S=S$ (the action is a number) and so
\beq 
\tilde\ph\tilde L\tilde\ph=\ph L\ph=
\tilde\ph e^{-\si}\:L\:e^{-\si}\tilde\ph\:.
\eeq
As a consequence
\beq 
\tilde\ph\tilde L\tilde\ph=\tilde\ph\ag
-\lap_{\tilde g}+\frac{\la}{2}\tilde\phi_c^2+\xi_D\tilde R
+e^{-2\si}[m^2+(\xi-\xi_D)R]\cg\tilde\ph\:,
\eeq
where $\tilde\phi_c$ is the classical solution (background field) 
and $\xi_D=(D-1)/4(D-2)$. 
As a result,  for a conformally coupled ($\xi=\xi_D$)
massless scalar field, the action in invariant in form. 

The first heat-kernel coefficients related to a generic operator 
of the form $-\lap-2W^k\nabla_k+M^2$ on a Riemannian, smooth manifold
without boundary read 
\beq 
&& a_0=1\:,\hs a_1=\frac{R}{6}-M^2-W^2-\hat\nabla_k W^k\:,\nn\\
&& a_2=\frac{a_1^2}{2}+\frac{\hat\lap a_1}{6}
+\frac{1}{180}(\lap R+R_{ijrs}R^{ijrs}-R_{ij}R^{ij})\:.
\eeq
Since in general $W^k$ could be a matrix, we have introduced the connection
$\hat\nabla_k f=\nabla_k+[W_k,f]$.

For the operator $L$ we are dealing with, $W_k=0$ and 
$M^2=\la\phi_0^2/2+m^2+\xi R$, while, for $\tilde L$, 
$\tilde M^2=\la\tilde\phi^2_0/2+\xi_D\tilde R+e^{-2\si}[m^2+(\xi-\xi_D)R]$.
Then one has
\beq 
&& \tilde a_1=\frac{\tilde R}{6}-\tilde M^2=
e^{-2\si}\aq a_1-\at\xi_D-\frac16\ct\Si\cq\:,\nn\\
&& \tilde a_2=\frac{\tilde a_1^2}{2}+\frac{\lap_{\tilde g}\tilde a_1}{6}+
+\frac{1}{180}(\lap_{\tilde g}\tilde R
+\tilde R_{ijrs}\tilde R^{ijrs}-\tilde R_{ij}\tilde R^{ij})\:.
\eeq
The relation between $\tilde a_2$ and $a_2$ in general is very complicated, 
but in the case of conformally coupled fields in $4$-dimensions.
In such a case we have in fact 
\beq 
\tilde a_1=e^{-2\si}a_1\:,\hs\hs
\tilde a_2=e^{-4\si}\at a_2-\frac{1}{3}\:\lap\si+\nabla_kV^k\ct\:,
\eeq
where $\nabla_kV^k$ is the total divergence due to  the 
geometric part of $a_2$. From the latter equation, the well known result
\beq 
\tilde A_2=\frac{1}{(4\pi)^2}\:\int\:d^4x\sqrt{\tilde g}\:\tilde a_2=
\frac{1}{(4\pi)^2}\:\int\:d^4x\sqrt{g}\:a_2=A_2\,,
\eeq
follows.
For the $P_n$ and $Q_n$ coefficients 
there are no general expressions in terms of geometrical 
quantities and so one cannot say anything about their 
conformal transformation properties, without to go to 
consider the specific problem.  

To conclude this Section, we present a simple derivation of
the conformal transformation properties of the one-loop effective action, 
which, according to generalised zeta-function regularisation, is given by
\beq 
W=-\log Z=S-\frac14\:\om''(0|L/\mu^2)\:.
\eeq
In general, this is not invariant even for conformally coupled fields due to 
the presence of the  functional measure, which breaks the symmetry. 
This phenomenon is well known and, in the physical literature, 
it is called ``conformal or trace anomaly''.

We set 
\beq 
&&\ln\tilde Z\equiv\ln Z[\tilde\phi,\tilde g]=J[g,\tilde g]\:\ln Z[\phi,g]
\nn\\
&&\ln J[g,\tilde g]=-(\tilde W-W)
\eeq
and consider a family of continuous transformations of the form
\beq 
g^q_{ij}=e^{2q\si}g_{ij}\:,\hs\hs g^0_{ij}=g_{ij}\:,
\hs\hs g^1_{ij}=\tilde g_{ij}\:.
\eeq
\beq 
\varphi_q=e^{q\si}\varphi\:,\hs\hs L_q=e^{-q\si}Le^{-q\si}\:,\hs\hs
W_q=W[\phi_q,g^q]\:. 
\eeq
For an infinitesimal transformation we get
\beq 
\ln J[g^q,g^{q+\de q}]=-(W_{q+\de q}-W_q)
=-\de\:S+\frac{\de\om''(0|L_q/\mu^2)}{4}\:.
\label{lJ1}
\eeq
We observe that 
\beq 
\de\om(s|L_q/\mu^2)=\de\aq s\Tr L_q^{-s}\cq=
2\de q\:\int\:d^Dx\sqrt{g^q}\:\si(x)s\:\om(s;x|L_q/\mu^2)\:,
\label{lJ2}\eeq
where $\om(s;x|L)$ represents the diagonal kernel of $\om(s|L)$, that is
\beq 
\om(s|L)=\int\:d^Dx\sqrt{g}\:\om(s;x|L)\:.
\eeq

In order to go on in arbitrary dimensions, we consider a more general
version of Eqs.~(\ref{tas0}) and (\ref{cit}), that is
\beq 
\Tr e^{-tL}\sim\sum_{0}^{\ii}\:B_jt^{j-D/2}
+\sum_{0}^{\ii}\:P_j\ln t\:t^{j-D/2}\:,
\label{kk0}\eeq 
\beq
\omega(s|L)=-P_{D/2}(L)+[B_{D/2}(L)-\ga P_{D/2}(L)]\:s+O(s^2)
\label{cit2}
\eeq
and suppose the local version
\beq
\omega(s;x|L)=-\frac{p_{D/2}(x|L)}{(4\pi)^{D/2}}
+\frac{[b_{D/2}(x|L)-\ga p_{D/2}(x|L)]\:s}{(4\pi)^{D/2}}+O(s^2)\:,
\label{citLoc}
\eeq
\beq 
B_n=\frac{1}{(4\pi)^{D/2}}\:\int\:d^Dx\sqrt{g}\:b_n(x|L)\:,\hs\hs
P_n=\frac{1}{(4\pi)^{D/2}}\:\int\:d^Dx\sqrt{g}\:p_n(x|L)\:,
\eeq
to be valid.
Then, from Eqs.~(\ref{lJ2}) and (\ref{lJ2}) we get
\beq 
\de\om''(s|L_q)=4\de q\:\int\:d^Dx\sqrt{g^q}\:\si(x)
[b_{D/2}(x|L_q/\mu^2)-\ga p_{D/2}(x|L_q/\mu^2)]\:.
\eeq 
Now, integrating (\ref{lJ1}) with respect to $q$ we obtain the
final formula
\beq 
\ln J[g,\tilde g]=
\frac{1}{(4\pi)^{D/2}}\int_0^1\:dq\int\:d^Dx\:\sqrt{g^q}\:\aq
b_{D/2}(x|L_q)-(\ga+\ln\mu^2)\:p_{D/2}(x|L_q)\cq\:.
\eeq
Here we have assumed the classical theory to be conformal invariant,
that is $\de S=0$. In such a case, when $p_{D/2}=0$,
the latter equation gives rise
to the well known form of the one-loop effective action.

\section{A non local interaction example}

In this Appendix we shall present another explicit example in which
logarithmic terms appear in the heat-kernel expansion.

We  consider the following toy model which, in a 
simplified manner, mimics an interacting  scalar field defined on 
a non-compact flat manifold with pairs of noncommuting coordinates.
Let the $D$-dimensional manifold be $M^D=\R^d\times\R^p$. Moreover,
let be $L_d$ and $L_p$ Laplacian like operators on $\R^d$ and $\R^p$ 
respectively with $D=p+d$ and $L_D=L_d+L_p$.
The model may be defined by the classical Euclidean action 
\begin{eqnarray}
S=\int dx \left[\frac{1}{2}\phi\left( L_D +\frac{a^2}{L_p}\right)\phi
+ V(\phi)\right]\,.
\label{f1}
\end{eqnarray}
The non-local interaction mimics the ``noncommuting'' manifold $\R^p$ as soon 
as $p$ is even and the 
parameter $a^2$ controls its presence. The self-interacting potential is given 
by 
\begin{eqnarray}
V(\phi)=\frac{\phi^r}{r!}\,.
\label{f11}
\end{eqnarray}
In renormalisable theories, the power $r$ is non arbitrary, but
it is related to the dimension. In fact one has the possible couples: 
$(D=3,r=6)$, $(D=4,r=4)$, $(D=6,r=3)$,...
The non local one-loop fluctuation operator reads
\begin{eqnarray}
L=L_D +\frac{a^2}{L_p}+ V''(\phi_c)=L_D +\frac{a^2}{L_p}+M^2\,,
\hs\hs M^2=\frac{\phi_c^{r-2}}{(r-2)!}\:.
\label{f2}
\end{eqnarray}
The heat-kernel trace and zeta function can be exactly evaluated and read
respectively
\begin{eqnarray}
\Tr e^{-tL}=\frac{2V_Da^{p/2}}{(4\pi)^{D/2}\Gamma(p/2)}
\frac{e^{-tM^2}\:K_{p/2}(2at)}{t^d/2}\,,
\end{eqnarray}
\begin{eqnarray}
\zeta(s|L)&=&
\frac{2^{p+1}\sqrt \pi\,V_Da^p}{(4\pi)^{D/2}\Gamma(p/2)(M^2-2a)^{s+(p-d)/2}}
\:\frac{\Gamma(s-(p-d)/2)\Gamma(s-D/2)}{\Gamma(s)\Gamma(s+(1-d)/2)}
\nonumber \\ 
&&\hs\times F\left(s+\frac{p-d}2,1+\frac p2;s+\frac{1-d}2,\frac{M^2-2a}{M^2+2a}\right)\,,
\label{f4}
\end{eqnarray}
$K_\nu(z)$ being the modified Bessel function and $V_D$ the volume of
the whole manifold and
$F(\alpha,\beta;\gamma,z)$ the hypergeometric function.

The function $\Tr e^{-tL}$ can be written as an exact series of $t$,
which assumes defferent forms for odd and even $p$. 
In fact for odd $p=n+1/2$ we have
\beq 
\Tr e^{-tL}=
\frac{\sqrt\pi\,V_D e^{-tM^2}e^{-2at}}{(4\pi t)^{D/2}\Ga(n+1/2)}
\sum_{k=0}^n\:\frac{(n+k)!\,(at)^{n-k}}{k!\,(n-k)!\,4^k}\:,
\eeq
while for even $p=2n$
\beq 
\Tr e^{-tL}&=&
\frac{V_D e^{-tM^2}}{(4\pi t)^{D/2}}\ag
\sum_{k=0}^{n-1}(-1)^k\:\frac{(n-k-1)!\,(at)^{2k}}{(n-1)!}
\cp\nn\\&&\ap -(-1)^n
\sum_{k=0}^{\ii}\:\frac{(at)^{2(n+k)}}{(n-1)!\,k!\,(n+k)!}
\aq 2\ln(at)-\psi(k+1)-\psi(n+k+1)\cq
\cg\:.
\eeq
One can see that logarithmic terms in the heat expansion only appear
for even $p$. Moreover, one has a pure logaritmic term only for 
even $d$,  equal or greather than $p$.
In particular, for $D=4$, $d=p=2$, $M^2=\lambda\frac{\phi_c^2}{2}$ 
one obtains
\begin{eqnarray}
\mbox{Tr} e^{-tL}&=&\frac{V_4a\:e^{-tM^2}}{8\pi^{2}t}\:
K_{1}(2at)\nn\\
&\simeq& \frac{V_4}{8\pi^2}
\aq \frac{1}{2 t^2} -\frac{M^2}{2 t}
+\frac{M^4}{4}+a^2(\ga+\ln a-1/2)+a^2 \ln t+O(t,t\ln t)\cq\,.
\end{eqnarray}

\end{document}